\documentclass[12pt]{article}
\usepackage{epsf}
\usepackage{overcite}

\begin{document}


\bibliographystyle{prsty}


\title {Converging toward a practical solution of the Holstein molecular crystal model}

\author{
A.~H.~Romero${^{1,3}}$, David W. Brown${^2}$ and Katja Lindenberg${^3}$ \\
${^1}$ Department of Physics,\\
University of California, San Diego, La Jolla, CA 92093-0354 \\
${^2}$ Institute for Nonlinear Science,\\
University of California, San Diego, La Jolla, CA 92093-0402 \\
${^3}$ Department of Chemistry and Biochemistry,\\
University of California, San Diego, La Jolla, CA 92093-0340
}

\date{\today}

\maketitle


\begin{abstract}
We present selected results for the Holstein molecular crystal model in one space dimension as determined by the Global-Local variational method, including complete polaron energy bands, ground state energies, and effective masses.
We juxtapose our results with specific comparable results of numerous other methodologies of current interest, including quantum Monte Carlo, cluster diagonalization, dynamical mean field theory, density matrix renormalization group, semiclassical analysis, weak-coupling perturbation theory, and strong-coupling perturbation theory.
Taken as a whole, these methodologies are mutually confirming and provide a comprehensive and quantitatively accurate description of polaron properties in essentially any regime.
In particular, this comparison confirms the Global-Local variational method as being highly accurate over a wide range of the polaron parameter space, from the non-adiabatic limit to the extremes of high adiabaticity, from weak coupling through intermediate coupling to strong coupling.
\end{abstract}


\titlepage


\section{Introduction}

Great progress has been made in the last few years toward achieving the practical solution of the Holstein molecular crystal model \cite{Holstein59a,Holstein59b}, a practical solution being one which yields a sufficiently complete spectrum of information to a sufficiently high accuracy to satisfy most purposes.
A number of different methodologies, each important for the particular strength they lend to the problem, are proving increasingly mutually consistent if not always straightforwardly convergent in their conclusions.
In this paper, we compare a suite of our own recent results
\cite{Zhao94b,Brown97b,Romero98b,Romero98c,Romero98}
obtained using the Global-Local variational method with comparable results of other contemporary approaches, including
quantum Monte Carlo (QMC),
\cite{DeRaedt83,DeRaedt84,DeRaedt97,Lagendijk85,McKenzie96,Kornilovitch97a,Kornilovitch97b}
cluster diagonalization,
\cite{Kongeter90,Ranninger92,Alexandrov94a,Alexandrov95,Kabanov97,Marsiglio93,Marsiglio95,Salkola95,Eiermann96,Wellein96,Wellein97a,Wellein97b,Fehske97a}
dynamical mean field theory (DMFT),
\cite{Ciuchi93,Ciuchi94,Ciuchi95,Ciuchi97,Freericks93,Freericks94}
density matrix renormalization group (DMRG),
\cite{White92,White93,Ostlund95,Jeckelmann98,Jeckelmann97b,Rommer97}
as well as both weak-coupling perturbation theory (WCPT) 
\cite{Jeckelmann97b,Migdal58}
and strong-coupling perturbation theory (SCPT) \cite{Tyablikov52,Lang63,Gogolin82,Marsiglio95,Stephan96,Capone97}.
Every such comparison we have been able to implement has shown the Global-Local results to be among the best currently available, and where a better result obtains by another method, the quantitative discrepancy involved tends to be satisfyingly small.
Thus, although it is not the case that our methods are necessarily the most accurate possible in every instance, our methods are valid over {\it large enough} a region of the polaron parameter space, and {\it accurate enough} over their region of validity to be sufficient to a wealth of practical purposes.

In several recent works (I \cite{Zhao97a} - II \cite{Zhao97b} - III \cite{Brown97b}) we have presented a sequence of increasingly refined variational approaches to the problem of determining the lowest polaron energy band and the associated energy-momentum eigenfunctions for the Holstein molecular crystal model in one space dimension.
The Global-Local method generalizes those of Toyozawa \cite{Zhao97b,Toyozawa61,Toyozawa63,Sumi73,Toyozawa80,Emin73,Schopka91} and Merrifield \cite{Zhao97a,Merrifield64} by including local electron-phonon correlation channels under-represented in the former and global electron-phonon correlation channels under-represented in the latter, both of which are of particular significance in the local structure of the polaron.
While in the present work we lean heavily on the results of Paper III because of its superior accuracy, this sequential approach has proven instrumental in motivating our present study and presaging some of our conclusions.

This paper is organized as follows:
In Section 2, we present the model and states upon which the present work is based, and set down notation.
In Section 3, we focus on the global ground state energy, displaying specific results according to our own method and comparing our results with those of other authors and certain approximate formulas.
In Section 4, we present some particular examples of complete polaron energy bands and compare Global-Local results with specific results obtained by the DMRG method and by direct cluster diagonalization.
In Section 5, we turn to the polaron effective mass, computing effective mass curves cutting swaths through the polaron parameter space in several regimes, and make specific comparisons of Global-Local results with those of a variety of competing approaches.
Conclusions are summarized in Section 6.

\section{Model, States, Method}

As our system Hamiltonian, we choose the traditional Holstein Hamiltonian \cite{Holstein59a,Holstein59b}.
\begin{equation}
\hat{H} = \hat{H}^{el} + \hat{H}^{ph} + \hat{H}^{el-ph} ~,
\end{equation}
\begin{equation}
\hat{H}^{el} = E \sum_n a_n^{\dagger} a_n  - J \sum_n a_n^{\dagger} ( a_{n+1} + a_{n-1} ) ~,
\end{equation}
\begin{equation}
\hat{H}^{ph} = \hbar \omega \sum_n b_n^{\dagger} b_n ~,
\end{equation}
\begin{equation}
\hat{H}^{el-ph} = - g \hbar \omega \sum_n a_n^{\dagger} a_n ( b_n^{\dagger} + b_n ) ~,
\end{equation}
in which $a_n^\dagger$ creates an electron in the rigid-lattice Wannier state at site $n$, and $b_n^\dagger$ creates a quantum of vibrational energy in the Einstein oscillator at site $n$.
We presume periodic boundary conditions on a one-dimensional lattice of $N$ sites.
The electron transfer integral between nearest-neighbor sites is denoted by $J$,
$\omega$ is the Einstein frequency, and $g$ is the dimensionless local coupling strength.
(Except where displayed for formulaic clarity, the reference energy $E$ is set to zero throughout.)
The lattice constant does not appear explicitly in this formulation provided wave vectors are measured relative to it, as will be our convention.
Two dimensionless control parameters can be constructed from the three principal Hamiltonian parameters in different ways.
One such non-dimensionalizing scheme involves selecting the phonon quantum $\hbar \omega$ as the unit of energy; in these terms, the natural dimensionless parameters are the remaining coefficients of the electronic hopping term ($J/ \hbar \omega$) and electron-phonon interaction term ($g$).
This scheme is particularly appropriate when considering dependences on $J$ and/or $g$ at fixed $\omega$, such as we shall be concerned with in most of this paper.
This scheme is not so convenient near the adiabatic limit, however, where both $J / \hbar \omega$ and $g$ diverge in a certain fixed relationship.
In the adiabatic regime, it is convenient to non-dimensionalize by selecting the electron half-bandwidth $2J$ as the unit of energy; in these terms, the natural dimensionless parameters are the remaining coefficients of the phonon energy ($\gamma = \hbar \omega /2J$) and the electron-phonon interaction term, expressed in the form $\sqrt{\lambda / \gamma }$, where $\lambda = g^2 \hbar \omega / 2J $.
(Here, we follow the convention of including the site coordination number $z$ in the definition of $\lambda$, such that $\lambda = g^2 \hbar \omega / z J $.)
The adiabatic limit is reached by allowing $\gamma$ to vanish at arbitrarily fixed $\lambda$.
We distinguish these two options as non-adiabatic and adiabatic scaling conventions, respectively.
Except where explicitly noted, we conform to the non-adiabatic scaling convention.

Our central interest in this paper is in the polaron energy band, computed as
\begin{equation}
E(\kappa) = \langle \Psi ( \kappa ) | \hat{H} | \Psi ( \kappa ) \rangle ~,
\end{equation}
wherein $\hat{H}$ is the total system Hamiltonian and $\kappa$ is the total {\it joint} crystal momentum label of the electron-phonon system.

All of our calculations are performed using normalized Bloch states
\begin{equation}
| \Psi ( \kappa ) \rangle = | \kappa \rangle / \langle \kappa | \kappa \rangle ^ {1/2} ~.
\end{equation}
These states are eigenfunctions of the appropriate total momentum operator and orthogonal for distinct $\kappa$, making variations for distinct $\kappa$ independent \cite{Lee53}.
Since selected $\kappa$ values such as the global polaron ground state ($\kappa = 0$) are generally insufficient to convey a very complete picture of polaron structure, we have found it important to examine {\it complete} variational solutions both for quality assurance and to extract the best description of polaron structure.
By ``complete'' variational solutions, we mean a set of $N$ variational energies $E(\kappa)$ and $N$ polaron Bloch states $| \Psi ( \kappa ) \rangle$; the latter being described by a distinct set of variational parameters for each $\kappa$.
The set of $E(\kappa)$ so produced constitute an estimate (upper bound) for the polaron energy band \cite{Lee53,Toyozawa61}.

The Global-Local method \cite{Brown97b} represents polaron structure through three classes of variational parameters $\{ \alpha_n^\kappa , \beta_q^\kappa , \gamma_q^\kappa \}$.
\begin{eqnarray}
|  \kappa \rangle  &=& \sum_{n n_a } e^{i \kappa n }
\alpha_{n_a -n}^\kappa a^{\dag}_{n_a} \exp \left\{ - N^{-\frac{1}{2}} \sum_q
[ ( \beta^\kappa_q e^{-iqn} - \gamma_q^\kappa e^{-iq n_a } ) b^{\dag}_q - H.c. ] \right\} |0\rangle ~.
\label{eq:gl}
\end{eqnarray}
We note here for later emphasis that a solution of the Global-Local method for any particular $\kappa$ is contained in the $3N$ complex quantities $\{ \alpha_n^\kappa , \beta_q^\kappa, \gamma_q^\kappa \}$, and that a {\it complete} solution is contained in $3N^2$ complex quantities, reducible to $O\{ \frac 3 2 N^2 \}$ independent real quantities utilizing Hamiltonian symmetries.
We typically computed complete energy band structures on $32$-site lattices for every parameter set $(J/ \hbar \omega , g)$ we considered; thus, the solution for any particular $\kappa$ is encoded in $96$ independent real numbers ($48$ for the ground state), and the complete solution (all $\kappa's$) in $1536$ independent real numbers (see, e.g., Figure~\ref{fig:g1j1} in Section 4).

Since our solutions are encoded in a relatively small number of variational parameters (compared to some other calculations we shall consider), even ``complete'' band structure solutions can be stored compactly, allowing a ``library'' of polaron band structures to be archived and revisited at leisure.

We solve the variational equations by relaxation techniques \cite{Zhao94b,Press92,Giordano97} through which an initializing state is iteratively refined toward the self-consistent target state.
Unlike many other methods, nearby solutions can be used to initialize new calculations, accelerating convergence and reducing the need to obtain new solutions from scratch.
Such initializing solutions might be obtained from nearby points in parameter space, for example, or might be lower-precision solutions obtained from prior calculations at the same parameters and $\kappa$.
A library of polaron structures sampling the polaron parameter space thus facilitates the acquisition of new solutions.

The computational time required for our calculation varies according to the region of the parameter space examined, the error tolerance required, and the ``scope'' of the calculation.
In the intermediate coupling regime, complete band structures computed to tolerances adequate for this paper take one to two hours on a single-processor Sun Microsystems Ultra Sparc I workstation; optimized to compute only the effective mass, only a few minutes per effective mass value are required.

More intensive calculation is required as one moves out from the intermediate coupling region, but in no reasonably general case for which we have achieved converged results has a complete set of $N$ band energies and $N$ Bloch states required more than two to four hours for one $(J/ \hbar \omega , g)$ point.
Some special cases presented greater difficulty; e.g., the limit of weak coupling for $J / \hbar \omega \sim 1/4$ proved inaccessible due to deteriorating numerical precision, and convergence in the vicinity of the self-trapping transition grew more challenging with increasing adiabaticity.

The self-consistency equations that follow from applying the variational principle to this class of trial states, the method of solving those equations, and sample results have been detailed in Paper III \cite{Brown97b}.
In the following sections, we compare polaron energy band characteristics as determined by the Global-Local method with a variety of alternative descriptions, but apart from the one complete solution illustrated in Figure~\ref{fig:g1j1}, we defer to a subsequent work \cite{Romero98} any detailed discussion of the particulars of internal polaron structure as reflected in the variational parameters themselves.

\section{Ground State Energy}

Without question, the single most important state in any quantum system is the ground state, and by association, the single most important energy is the ground state energy.
That being said, we emphasize quickly that a particular numerical value of the ground state energy {\it in itself} contains very little information about the system that is of practical use.
For most purposes, we require relationships {\it between} energies; e.g., effective masses, bandwidths, densities of states.
For such purposes we require multiple energies, all but one of which are {\it excited states} in the global sense.
Variational approaches such as ours divide the total space of the problem into subspaces of the total crystal momentum within which the ground state of each $\kappa$ sector is determined independently.
In this sense, the {\it global} ground state energy $E(0)$ is only one of $N$ independent $\kappa$-sector ground state energies $E(\kappa)$ computed on an equal footing.
There is no guarantee that the numerical value of $E(0)$ constitutes any more (or less) accurate an estimate of its particular target value than any other $E( \kappa )$ so determined.

Nonetheless, it is the global ground state energy $E(0)$ that is the most common denominator of diverse variational approaches, including those based on localized states or other states that may not be well adapted to the properties expected of energy-momentum eigenfunctions.
Thus, we have compiled in Tables~\ref{tab:grounda} and \ref{tab:groundb} a number of ground state energies from our own calculations together with a sampling of others for two particular sets of parameters of the Holstein Hamiltonian.


\begin{table*}[t]
\centering
\caption{Ground State Energies as computed by various methods for $(J/ \hbar \omega , g) = (1,1)$, $( \gamma , \lambda ) = ( \frac 1 2 , \frac 1 2 )$.
The table is broken at the most likely location of the true ground state energy.
Values indicated by dagger symbols are obtained from limiting formulae that at these parameter values are clearly beyond their limits of validity.}
\vskip 2mm
\begin{tabular}{|l|l|l|} \hline
Value $[ \hbar \omega ]$ & Type of Method & Source \\ \hline \hline
- 1.73575 $\dagger$ & Non-adiabatic small polaron & Holstein \cite{Holstein59b}, Eq. \ref{eq:smpolgrnd}\\ \hline
- 1.97 & QMC $N = 32$ $\beta = 1$ $m = 10$ & De Raedt and Lagendijk \cite{DeRaedt83,DeRaedt97} \\ \hline
- 2.00000 $\dagger$ & Adiabatic strong coupling & Gogolin \cite{Gogolin82}, Eq. \ref{eq:gogolin}\\ \hline
- 2.08614 & Semiclassical variation & Kalosakas, Aubry, and Tsironis \cite{Kalosakas97a,Kalosakas97b} \\ \hline
- 2.35 & QMC $N = 32$ $\beta = 5$ $m = 32$ & De Raedt and Lagendijk \cite{DeRaedt83,DeRaedt97} \\ \hline
- 2.40 & Dynamical Mean Field & Ciuchi, et al. \cite{Ciuchi97} \\ \hline
- 2.44721 & $2^{nd}$ order WCPT & Eq. \ref{eq:wcptgrnd} \\ \hline
- 2.45611 & Merrifield variation & Zhao, Brown, and Lindenberg \cite{Zhao97a} \\ \hline
- 2.46 & QMC $N = 32$ $\beta = 20$ $m = 256$ & De Raedt \cite{DeRaedt97} \\ \hline
- 2.46869 & Toyozawa variation & Zhao, Brown, and Lindenberg \cite{Zhao97b} \\ \hline
- 2.46931 & Global-Local variation & Figure~\ref{fig:grndj}, this paper \\ \hline
- 2.46968 & DMRG $N = 32$ & Jeckelmann and White \cite{Jeckelmann98,Jeckelmann97b} \\ \hline \hline
- 2.471 & Cluster diagonalization, $N = 6$ & Alexandrov, et al. \cite{Alexandrov94a,Kabanov97,footnote1} \\ \hline
- 3.08959 $\dagger$ & $2^{nd}$ order SCPT & Marsiglio \cite{Marsiglio95}, Stephan \cite{Stephan96}, Eq. \ref{eq:scptgrnd}\\ \hline
\end{tabular}
\label{tab:grounda}
\end{table*}

\begin{table*}
\centering
\caption{Ground State Energies as computed by various methods for $(J/ \hbar \omega , g) = (1,\sqrt{2})$, $( \gamma , \lambda ) = ( \frac 1 2 , 1 )$.
The table is broken at the most likely location of the true ground state energy.
Values indicated by dagger symbols are obtained from limiting formulae that at these parameter values are clearly beyond their limits of validity.}
\vskip 2mm
\begin{tabular}{|l|l|l|} \hline
Value $[ \hbar \omega ]$ & Type of Method & Source \\ \hline \hline
- 2.27067 $\dagger$ & Non-adiabatic small polaron & Holstein \cite{Holstein59b}, Eq. \ref{eq:smpolgrnd}\\ \hline
- 2.50000 $\dagger$ & Adiabatic strong coupling & Gogolin \cite{Gogolin82}, Eq. \ref{eq:gogolin}\\ \hline
- 2.51815 & Semiclassical variation & Kalosakas, Aubry, and Tsironis \cite{Kalosakas97a,Kalosakas97b} \\ \hline
- 2.73 & QMC $N = 32$ $\beta = 1$ $m = 10$ & De Raedt and Lagendijk \cite{DeRaedt83,DeRaedt97} \\ \hline
- 2.86 & QMC $N = 32$ $\beta = 5$ $m = 32$ & De Raedt and Lagendijk \cite{DeRaedt83,DeRaedt97} \\ \hline
- 2.89 & Dynamical Mean Field & Ciuchi, et al. \cite{Ciuchi97} \\ \hline
- 2.89442 & $2^{nd}$ order WCPT & Eq. \ref{eq:wcptgrnd} \\ \hline
- 2.93301 & Merrifield variation & this paper \\ \hline
- 2.99172 & Toyozawa variation & this paper \\ \hline
- 2.99802 & Global-Local variation & Figure~\ref{fig:grndj}, this paper \\ \hline
- 2.99883 & DMRG $N = 32$ & Jeckelmann and White \cite{Jeckelmann98,Jeckelmann97b} \\ \hline \hline
- 3.000 & Cluster diagonalization, $N = 6$ & Alexandrov, et al. \cite{Alexandrov94a,Kabanov97,footnote1} \\ \hline
- 3.00 & QMC $N = 32$ $\beta = 20$ $m = 256$ & De Raedt \cite{DeRaedt97} \\ \hline
- 3.05279 $\dagger$ & $2^{nd}$ order SCPT & Marsiglio \cite{Marsiglio95}, Stephan \cite{Stephan96}, Eq. \ref{eq:scptgrnd}\\ \hline
\end{tabular}
\label{tab:groundb}
\end{table*}

We have chosen the parameter sets $(J/ \hbar \omega , g) = (1,1)$ and $(1, \sqrt{2})$ largely because these points are commonly transected by the ground state curves of different approaches in the literature and thus permit the widest possible comparisons to be made.
We have also chosen these parameters values because they are {\it moderate}, falling in a regime where no particular limiting theory is favored.
Indeed, several results included in our tabulations (indicated by dagger symbols) are well-known asymptotic results not strictly valid in these intermediate scenarios.

In making the selection of data to display in Tables~\ref{tab:grounda} and \ref{tab:groundb}, we have not attempted to be exhaustive, but to present a sampling of results from differing methodologies, highlighting more recent examples.
With the exception of the DMFT values, all tabulated data have been computed directly by us, obtained from cited analytic formulae, or obtained directly from the original authors by private communication.
The values shown for dynamical mean field theory have been measured from published figures, and are believed by us to fairly represent the published data to the number of significant digits displayed.

Since not all of the approaches sampled in Tables \ref{tab:grounda} and \ref{tab:groundb} are variational calculations, we should note that a comparison of a variational energy $E_{var}$ with a lower non-variational energy $E_{non}$ ($E_{var} > E_{non}$) in itself yields no conclusion regarding which value is ``better'' unless there exists independent proof that the non-variational energy lies {\it above} the ground state.
In the opposite circumstance ($E_{non} > E_{var}$), however, the variational principle alone assures that the variational energy is the better estimate.

The most probable value of the true ground state energy can be sorted out from the data in Tables \ref{tab:grounda} and \ref{tab:groundb} as follows:
In view of the fact that the uncertainties in the best QMC values tabulated are of the order of one percent or greater \cite{DeRaedt97}, the QMC values cannot be viewed as any more definitive than the Global-Local, DMRG, and cluster values that are all determined to higher precision and fall within the QMC uncertainties.
Further, since the density matrix renormalization group method has been shown to have a variational realization \cite{White92,White93,Ostlund95,Jeckelmann98,Jeckelmann97b,Rommer97}, it would appear that the true ground state energy should lie below the DMRG values.
On the other hand, as is discussed in greater detail in Section 4, cluster energies appear to converge {\it upward} with increasing cluster size, suggesting that the true bulk ground state energy lies {\it above} the $N=6$ cluster values.
Thus, each of the tables shown has been broken at the point where the best apparent upper bound to the ground state energy meets the best apparent lower bound.

\begin{figure}[htb]
\begin{center}
\leavevmode
\epsfxsize = 3.6in
\epsffile{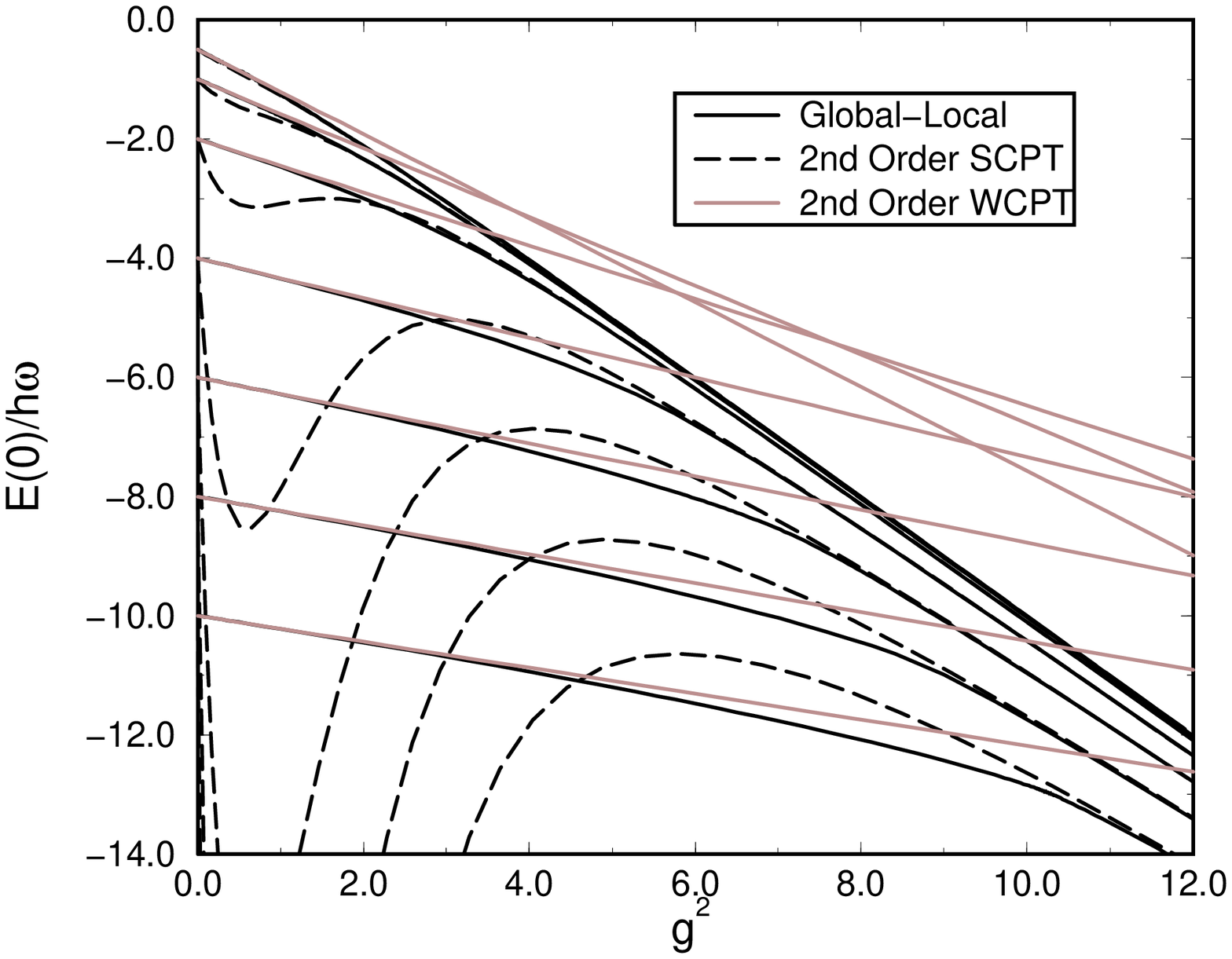}
\end{center}
\caption
{
The global ground state energy $E(0)$ vs. $g^2$ for $J/ \hbar \omega = 1/4$, $1/2$, $1$, $2$, $3$, $4$, and $5$ in order from top to bottom (note that $\lim_{g \rightarrow 0} E(0) = -2J$).
The straight dotted lines asymptoting each Global-Local curve at small $g^2$ are given by the second-order WCPT formula (\ref{eq:wcptgrnd}).
The arched dashed lines asymptoting each Global-Local curve at large $g^2$ are given by the second-order SCPT formula (\ref{eq:gogolin}).
}
\label{fig:grndj}
\end{figure} 

Turning our attention to the broader landscape of the system parameter space, Figure~\ref{fig:grndj} shows the dependence of the global ground state energy $E(0)$ on the coupling strength for assorted values of $J/ \hbar \omega$ from $1/4$ to $5$.
The overall behavior of the ground state energy is to trend between two asymptotic $g^2$ dependences with differing coefficients and offsets.
The crossover between these asymptotic trends occurs at the self trapping transition ($g \approx g_{ST}$), which through analyses presented elsewhere \cite{Romero98b,Romero98c,Romero98} has been found to be accurately located by the relation $g_{ST} = 1+\sqrt{J/ \hbar \omega}$.

Using weak-coupling perturbation theory, one can show that the leading dependence of the ground state energy on the coupling constant is given for any $J$ by
\begin{equation}
E(0) \approx - 2J - \frac {g^2 \hbar \omega} {\sqrt{ 1+4J / \hbar \omega }} ~.
\label{eq:wcptgrnd}
\end{equation}
This is, in fact, what we find to within numerical precision within the weak coupling regime of the Global-Local method (see Figure~\ref{fig:grndj}).

The weak-coupling estimate (\ref{eq:wcptgrnd}) is superior to the non-adiabatic small polaron estimate \cite{Tyablikov52,Holstein59b,Lang63}
\begin{equation}
E(0) \approx -g^2 \hbar \omega - 2J e^{-g^2}
\label{eq:smpolgrnd}
\end{equation}
for most $J$ provided the coupling strength is not too large ($g < g_{ST}$), though the two approximations agree when both $J/ \hbar \omega$ and $g$ are small.
When extrapolated into the adiabatic regime, however, (\ref{eq:smpolgrnd}) is not even qualitatively sensible at weak coupling for $J / \hbar \omega > 1/2$, and does not even approach the weak-coupling estimate until $g \approx g_{ST}$.
At strong coupling ($g \gg g_{ST}$), (9) approaches $-g^2 \hbar \omega$ for any $J$, which generally differs significantly from the true value of the ground state energy.
Thus, it appears that there is {\it nowhere} in the adiabatic regime where the non-adiabatic small polaron approximation provides a meaningful estimate of the ground state energy.

On the other hand, the adiabatic strong-coupling perturbation result (in one dimension) \cite{Gogolin82}
\begin{equation}
E(0) \approx -2J \lambda \left( 1 + \frac 1 {4 \lambda^2} \right) = - g^2 \hbar \omega - \frac {J^2} {g^2 \hbar \omega}
\label{eq:gogolin}
\end{equation}
correctly describes the asymptotic behavior at strong coupling for $g \gg g_{ST}$.
Both this adiabatic result and the non-adiabatic result (\ref{eq:smpolgrnd}) can be recovered from the second-order strong-coupling perturbation result \cite{Marsiglio95,Stephan96}
\begin{eqnarray}
E(0) &\approx& -g^2 \hbar \omega - 2J e^{-g^2} \nonumber \\
& & - \frac 2 {\hbar \omega} J^2 e^{-2g^2} [ f(2g^2) + f(g^2) ] ~,
\label{eq:scptgrnd}
\end{eqnarray}
\begin{equation}
f(x)= {\rm Ei}(x)-\gamma- \ln(x) ~,
\label{eq:fofx}
\end{equation}
in which ${\rm Ei}(x)$ is the exponential integral and $\gamma$ is the Euler constant.
Examples of this strong-coupling result are plotted in Figure~\ref{fig:grndj} together with comparable Global-Local and weak-coupling results.
The second order SCPT result is good for all $g$ provided $J/ \hbar \omega < 1/4$, and for all $J$ provided $g \gg g_{ST}$, but breaks down rather dramatically otherwise.

Thus, while both weak- and strong-coupling perturbation theories are clearly limited in scope, both are in quantitative agreement with results of the Global-Local method in the appropriate regimes, suggesting that a quite complete picture is available in the results of the Global-Local method augmented by the leading-orders of perturbation theory.
Indeed, this blended approach proves to be broadly useful, as is borne out in results presented elsewhere \cite{Romero98b,Romero98c,Romero98}.

\section{Energy Bands}

The archetypical result of polaron theory in the strong coupling limit is the band form \cite{Tyablikov52,Holstein59b,Lang63,Marsiglio95,Stephan96}
\begin{equation}
E(\kappa) = (E - 2J) - E_{B} - \frac 1 2 B [1 - \cos (\kappa) ] ~.
\end{equation}
The detailed dependence of the binding energy $E_{B}$ and bandwidth $B$ on $J$ and $g$ depend on regime, and it is through this dependence that changes in polaron structure are manifested in this limiting result.
The binding energy and bandwidth are related quite simply in the non-adiabatic regime ($E_{B} \sim g^2 \hbar \omega$, $B \sim 4J e^{-g^2}$), and less simply in the adiabatic regime.
When this band form is valid, the polaron is heavily dressed by phonons and the polaron bandwidth $B$ is small relative to {\it both} the bare electron bandwidth $4J$ and the bare phonon energy $\hbar \omega$; i.e., $B \ll Min \{4J,~\hbar \omega \}$.
This narrowing of the polaron band and the related increase in the effective mass are commonly characterized as aspects of polaron band {\it distortion}.

More generally, however, away from the strong coupling limit, polaron energy bands are {\it not} simple narrowed and shifted replicas of the bare band, but are more strongly distorted shapes whose non-sinusoidal dependence on $\kappa$ is a crucial reflection of polaron structure.
For such bands, the polaron binding energy, effective mass, and polaron bandwidth no longer stand in any simple relationship to each other.

\begin{figure}[htb]
\begin{center}
\leavevmode
\epsfxsize = 3.6in
\epsffile{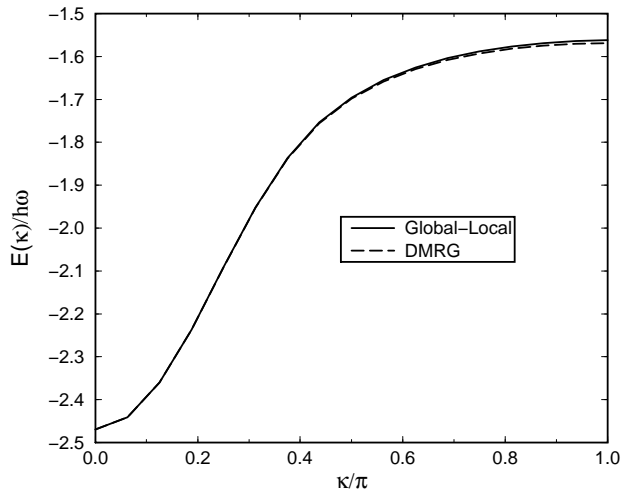}
\end{center}
\caption
{Comparison of Global-Local (GL) energy band results with comparable results of the density matrix renormalization group (DMRG) method for $J/ \hbar \omega = 1$, $g=1$.
$\kappa_c = \pi / 3$.
DMRG data kindly provided by E. Jeckelmann. \protect \cite{Jeckelmann97b}\protect .}
\label{fig:glvsdmrg}
\end{figure}

In the limit $g \rightarrow 0^+$ of the adiabatic regime ($J/ \hbar \omega > 1/4$), one finds that the polaron energy band assumes a {\it clipped} form,
\begin{eqnarray}
E( \kappa ) &= & (E - 2J) + 2J [ 1 - \cos ( \kappa ) ]  ~~~~~ | \kappa | < \kappa_c ~, \\
& = & (E - 2J) + \hbar \omega ~~~~~ ~~~~~ ~~~~~ ~~~~~  | \kappa | > \kappa_c ~,
\end{eqnarray}
reflecting the difference in the character of polaron states above and below the wave vector $\kappa_c$ (given by $2J [1 - \cos ( \kappa_c ) ] = \hbar \omega $) at which the bare electron energy band crosses into the one-phonon continuum.
Although this crisply clipped band form is strictly valid only in the limit $g \rightarrow 0^+$, its essential characteristics persist to non-trivial values of the electron-phonon coupling strength.
Figures~\ref{fig:glvsdmrg} and \ref{fig:welleinband}, for example, demonstrate the persistence of the clipped band form and of $\kappa_c$ as the relevant scale parameter for $g$ of order unity.

Figure~\ref{fig:glvsdmrg} compares a complete Global-Local energy band with a comparable energy band as computed independently by the DMRG method \cite{Jeckelmann98,Jeckelmann97b} for the case $J/ \hbar \omega =1$, $g=1$.
As was the case in the ground state energy comparisons of Tables \ref{tab:grounda} and \ref{tab:groundb}, the DMRG energies are slightly lower than the GL energies, the difference ranging from about $0.015\%$ at $\kappa = 0$ to about $0.46\%$ at $\kappa = \pi$; no fitting or numerical adjustment of any kind was made to normalize these two results.

\begin{figure*}[t]
\begin{center}
\leavevmode
\epsfxsize = 3.2in
\epsffile{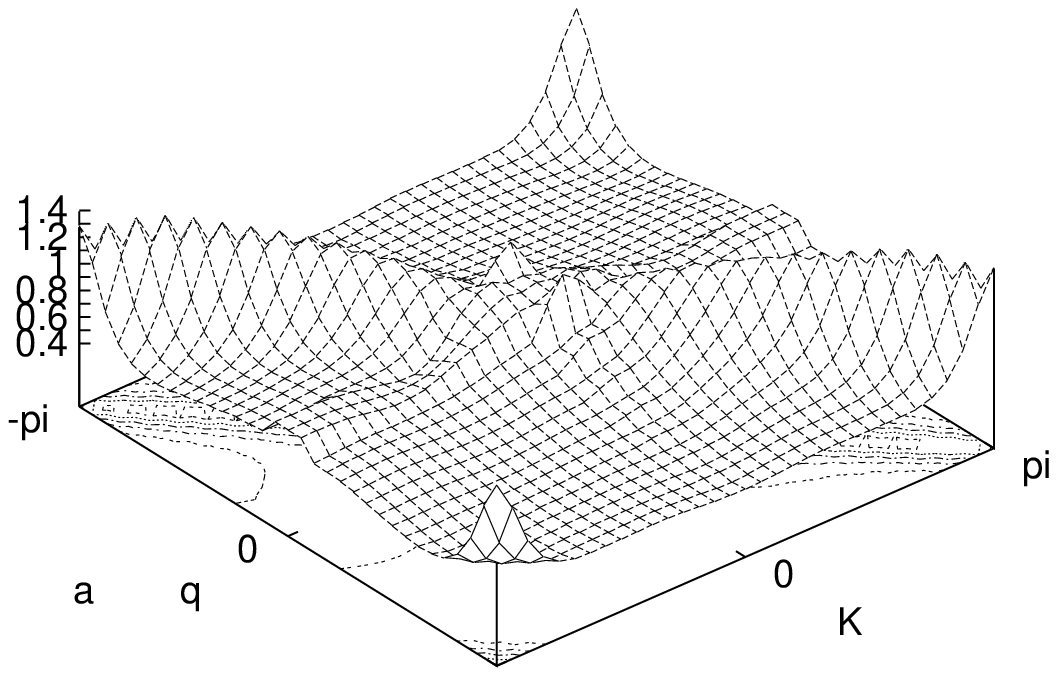}
\vspace{.1in}
\leavevmode
\epsfxsize = 3.2in
\epsffile{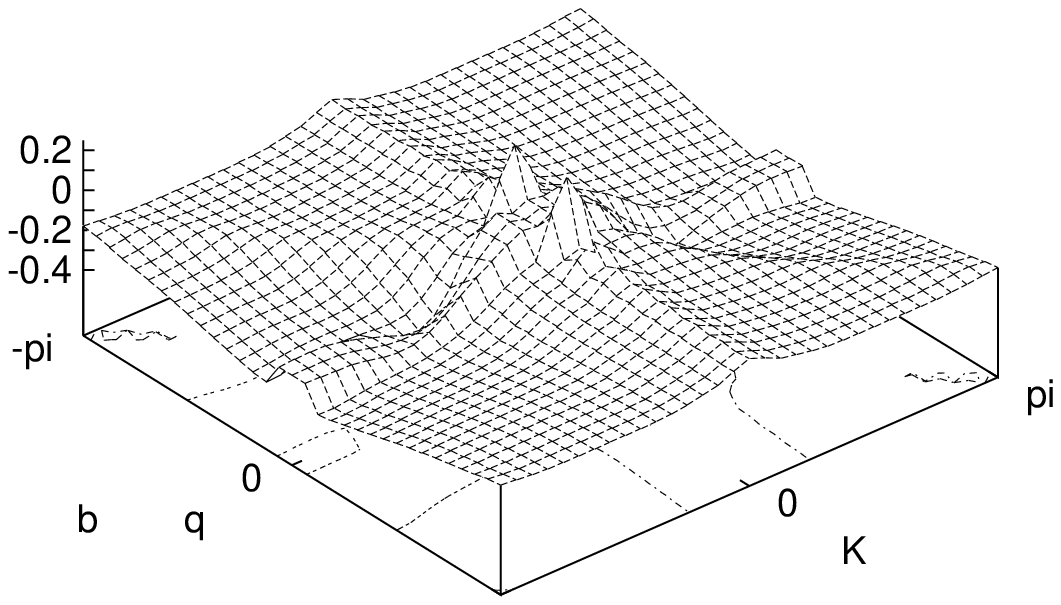}
\end{center}
\begin{center}
\leavevmode
\epsfxsize = 3.2in
\epsffile{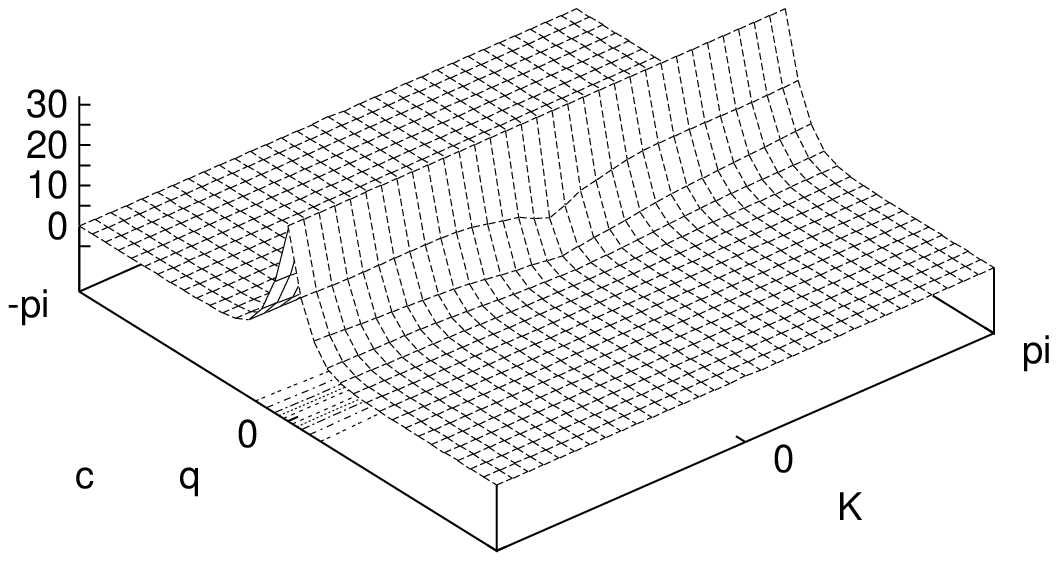}
\vspace{.1in}
\leavevmode
\epsfxsize = 3.2in
\epsffile{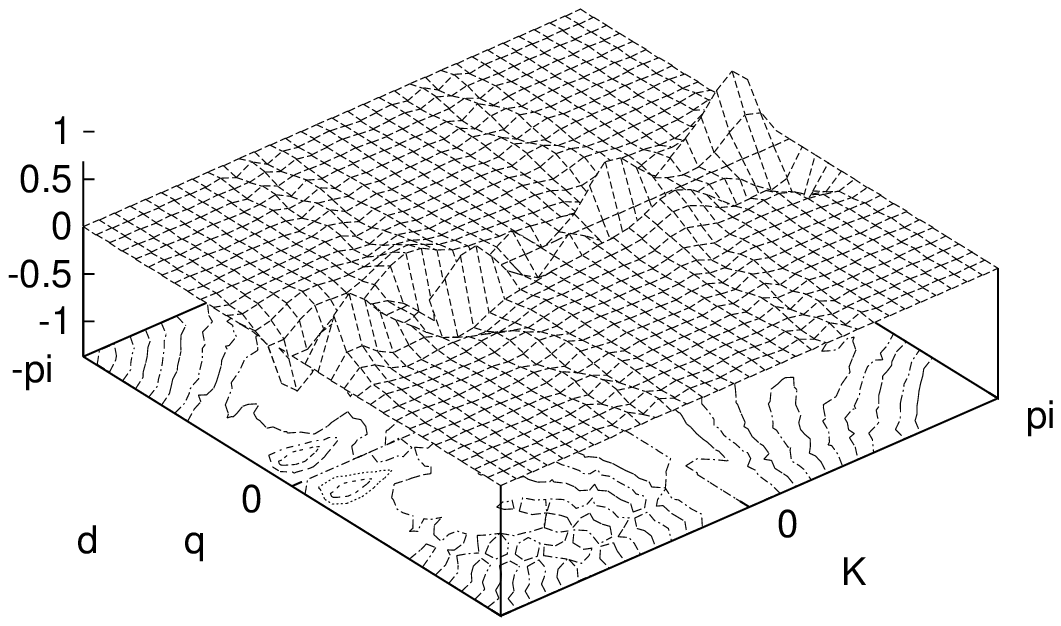}
\end{center}
\caption
{ Complete Global-Local solutions for the case $J / \hbar \omega =1$, $g=1$ corresponding to the ground state energy tabulation in Table~\ref{tab:grounda} and energy band presentation in Figure~\ref{fig:glvsdmrg}; a) $\beta_q^\kappa$, b) $\gamma_q^\kappa$, c) $Re \{ \alpha_n^\kappa \}$, d) $Im \{ \alpha_n^\kappa \}$.
$\kappa_c = \pi / 3$.}
\label{fig:g1j1}
\vspace{0.2in}
\end{figure*}

\begin{figure}[htb]
\begin{center}
\leavevmode
\epsfxsize = 3.6in
\epsffile{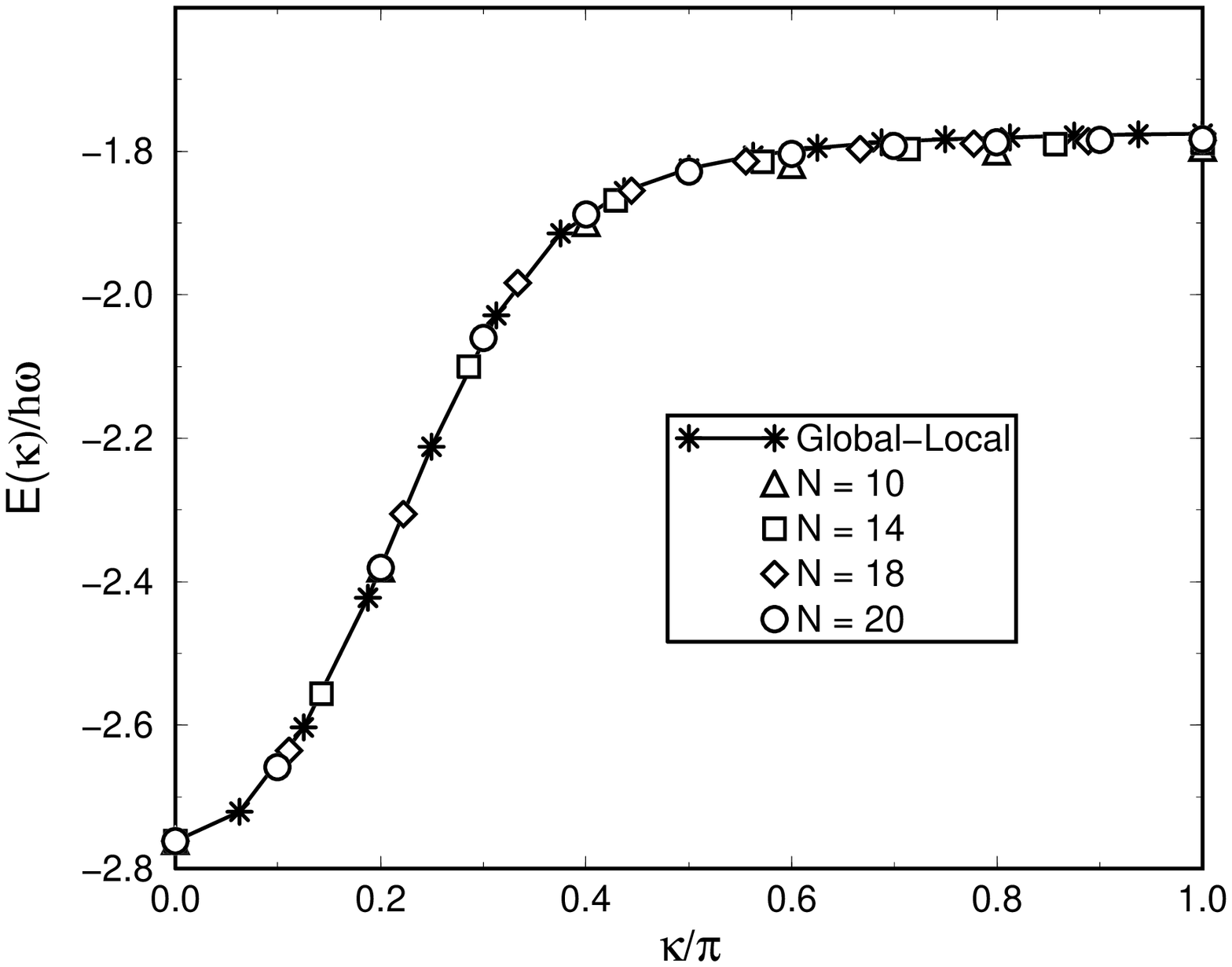}
\end{center}
\caption
{Comparison of our variational energy band with results of cluster diagonalizations for cluster sizes $N=10$, $14$, $18$, and $20$; $J / \hbar \omega  = 1.25$ and $g= \sqrt{0.5 \cdot 1.25} = 0.79 ...$, corresponding to $(\gamma , \lambda) = ( 0.4 , 0.25 )$.
$\kappa_c \approx 0.2952 \pi$.
Cluster data kindly provided by G. Wellein.\protect  \cite{Wellein97b}\protect .}
\label{fig:welleinband}
\end{figure}

The several surfaces completely detailing the electron-phonon structure underlying this energy band are shown in Figure~\ref{fig:g1j1}.

These surfaces show polaron structure to be composed of multiple distinguishable and characteristic features and correlations.
It is not our purpose here to analyze this internal structure in depth; \cite{Brown97b,Zhao97a,Zhao97b,Brown95a,Brown95b,Brown97a,Lindenberg97} however, a few outstanding features are worth noting.
First, in all components of the solution, there exist distinguishable characters in the inner Brillouin zone ($| \kappa | < \kappa_c$) and the outer Brillouin zone ($ | \kappa | > \kappa_c $).
These distinct characters are least evident in the electron amplitudes $\alpha_q^\kappa$, and most evident in the primary phonon amplitude $\beta_q^\kappa$.
In the inner zone, phonon structure consists of phonon amplitudes correlated with the electronic component in a manner that is structured, but largely local in character.
In the outer zone, phonon structure continues to have a localized component, but this component is dominated by a strong, momentum-rich, largely-delocalized component reflecting the strong influence of the one-phonon continuum on ``clipped'' polaron energy bands.
Although these inner and outer zone features become muted with increasing coupling strength and become nearly indistinguishable above the self-trapping transition, their qualitative character is pervasive.

\begin{figure}[htb]
\begin{center}
\leavevmode
\epsfxsize = 3.6in
\epsffile{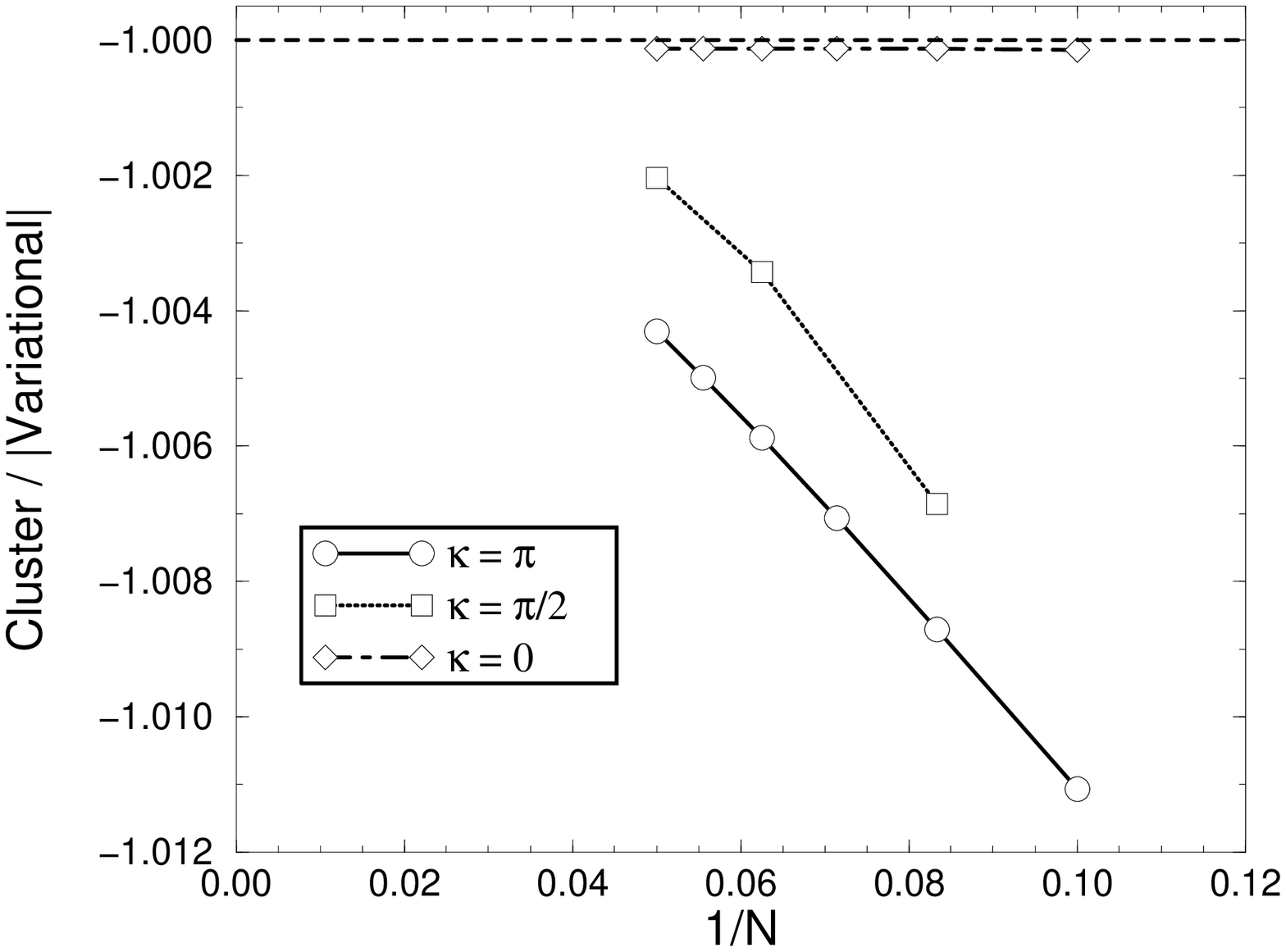}
\end{center}
\caption
{
Cluster energies for $\kappa = 0$, $\pi /2$, and $\pi$ and cluster sizes $N = 10$ to $20$ for $J / \hbar \omega  = 1.25$ and $g = \sqrt{ 0.5 \cdot 1.25} \approx 0.79 ...$.
Each cluster energy is divided by the absolute value of our variational energy appropriate to that particular $\kappa$, so that $-1$ indicates identical results under the two methods.  Cluster data kindly provided by G. Wellein. \protect \cite{Wellein97b}\protect
}
\label{fig:wellein}
\end{figure}

Figure~\ref{fig:welleinband} overlays the appropriate variational energy band computed by the Global-Local method upon data from several finite-cluster diagonalization results for the particular case $J / \hbar \omega =1.25$ and $g = \sqrt{ 0.5 \cdot 1.25} \approx 0.79 ...$.
No fitting has been performed to achieve the impressive degree of agreement evident in these independent results.
The variational results and the $N=20$ cluster results agree to multiple significant digits across the entire Brillouin zone, the relative difference being approximately 0.01\% at $\kappa=0$ and 0.4\% at $\kappa=\pi$.
This degree of agreement is actually even {\it better} than it appears at first glance, as may be inferred from the convergence trends in the cluster data:

Unlike variational bounds that converge {\it downward} toward the exact target energy, the trends evident in the data of Wellein and Fehske \cite{Wellein97a,Wellein97b} suggest that cluster energies converge {\it upward} with increasing cluster size.
Figure~\ref{fig:wellein} displays the ratio $E_N^{M=10} (\kappa) / | E_{N=32}^{GL} (\kappa) |$ for several $\kappa$ and $N$, where $E_N^M (\kappa)$ is the band energy determined by Lanczos diagonalization on a cluster of $N$ sites allowing up to $M$ total phonons, and $E_{N=32}^{GL} (\kappa)$ is the Global-Local variational results computed for 32 sites as throughout this paper.
The $\kappa = 0$ cluster energy is clearly well converged to a value slightly lower than, but nearly identical to the variational energy.
(That the $\kappa= 0$ cluster energy is nearly independent of $N$ suggests that the $N=6$ cluster energies \cite{Alexandrov94a,Kabanov97} included in Tables \ref{tab:grounda} and \ref{tab:groundb} are probably also well-converged.)
Cluster energies are less well-converged away from the Brillouin zone center, where the evident trends suggest that the bulk limits of the finite-$\kappa$ energies lie near the variational value.
Thus, though in principle and through sufficiently comprehensive computation on sufficiently large clusters, cluster diagonalization methods should best our variational approach at any $\kappa$, it appears that cluster sizes and retained phonon numbers must be significantly larger than those attempted to date in order improve significantly upon the Global-Local results for the polaron energy band {\it as a whole}.

One conclusion that can be drawn from these comparisons (considering, for example, a fixed $N$) is that the outer Brillouin zone ($| \kappa | > \kappa_c$) displays a greater sensitivity to contributions from higher phonon numbers than does the inner zone ($|\kappa | < \kappa_c$).
This is exactly what is to be expected from the $\kappa$-dependence of the mean phonon number, which is typically considerably smaller in the inner zone than in the outer zone (see, e.g., Figure 1 of Paper II \cite{Zhao97b}).

Owing to ultimate limitations of computing time and physical limitations of data storage, cluster diagonalizations generally are limited by the maximum dimension of the truncated Hilbert space that can be addressed by a particular computer.
For a one-electron problem on a lattice of $N$ sites, the dimension of the electronic subpace is $D_{el} = N$ and the dimension $D_{ph}$ of a truncated phonon subspace containing at most $M$ phonons is given by $D_{ph} = (N+M)!/ N!M!$.
Calculations have been reported for various combinations of $N$ and $M$ from $N=2$ to $N=20$ and $M$ up to $50$, each such calculation striking a unique balance between $N$ and $M$ consistent with computational resources.
Table~\ref{tab:cluster} presents values conveying the scale of the diagonalization problem for clusters sizes and phonon numbers up to $N = M = 32$.
Clearly, the scenarios represented by the lower right half of the tabulated values lie beyond the scope of machine diagonalization.

\begin{table*}[t]
\centering
\caption{Cluster Facts}
\vskip 2mm
\begin{tabular}{|r|c|c|r|r|r|} \hline
Cluster & $J_{min} / \hbar \omega $ & $J_{max} / \hbar \omega $ & $D_{tot}, M=10$ & $D_{tot}, M=16$ & $D_{tot}, M=32$ \\ \hline \hline
2 sites & no value & no value & $132$ & $306$ & $1122$ \\ \hline
4 sites & $0.5000$ & $0.500$ & $4.0 \times 10^3$ & $1.9 \times 10^4$ & $2.4 \times 10^5$ \\ \hline
6 sites & $0.3333$ & $1.000$ & $4.8 \times 10^4$ & $4.5 \times 10^5$ & $1.7 \times 10^7$ \\ \hline
10 sites & $0.2764$ & $2.618$ & $1.8 \times 10^6$ & $5.3 \times 10^7$ & $1.5 \times 10^{10}$ \\ \hline
16 sites & $0.2599$ & $6.569$ & $8.5 \times 10^7$ & $9.6 \times 10^9$ & $3.6 \times 10^{13}$ \\ \hline
20 sites & $0.2563$ & $10.22$ & $6.0 \times 10^8$ & $1.5 \times 10^{11}$ & $2.5 \times 10^{15}$ \\ \hline
32 sites & $0.2524$ & $26.02$ & $4.7 \times 10^{10}$ & $7.2 \times 10^{13}$ & $5.9 \times 10^{19}$ \\ \hline
\end{tabular}
\label{tab:cluster}
\end{table*}

Implementation of the density matrix renormalization group method also involves a truncation of the total quantum Hilbert space; however, the limitations imposed by such truncations appear to be less onerous than in the case of cluster diagonalization, permitting relatively large calculations to be accomplished in reasonable time on desktop platforms, much as the global local method.

By contrast, the ``size'' of a Global-Local variation does not scale directly with phonon number, since the nature of the Global-Local approximation is not to truncate the phonon Hilbert space but to characterize the phonon distribution for $M=\infty$ through a relatively small number of parameters.
For example, in all the calculations presented in this paper, only $3N$ independent variational parameters were required to flexibly represent the polaron state to high precision at any $\kappa$ ($96$ for $ | \kappa | \in (0, \pi)$, $48$ for $ | \kappa | = 0 , \pi$) with {\it no} restriction on phonon numbers, vastly fewer than comparable cluster diagonalizations.

It is also the case that the Global-Local method is not at its best for small clusters, but improves in quality with increasing $N$ (though not without limit).
This improvement is realized because the total phonon state is represented by a superposition of $N$ distinct phonon coherent states; such a superposition becomes more flexible as the number of coherent states in the superposition increases.

Our own comparisons with 6-cluster ground state energies in Tables \ref{tab:grounda} and \ref{tab:groundb} proved mutually favorable, as has our comparison with the $N= 10 - 20$ cluster ground states as computed by Wellein and Fehske \cite{Wellein97a,Wellein97b}.
However, we are more generally concerned with the {\it complete} band structure of the polaron, which is not well captured by calculations on small clusters as can be seen from the following:

From weak-coupling nearly up to the self-trapping transition, polaron bands exhibit distinct character for wave vectors above and below $\kappa_c$ that must be resolved if one is to faithfully represent polaron structure.
This underscores the importance in {\it any} finite-lattice calculation of maintaining the lattice size $N$ in relation to the tunneling matrix element $J$ such that at least one finite-$\kappa$ point is sampled on both the high and low sides of $\kappa_c$.
For a given $N$, this implies that only for $J \in ( J_{min} , J_{max} )$ can {\it both} the inner and outer polaron band structure be characterized independently,
where
\begin{eqnarray}
J_{max} &=& \frac {\hbar \omega} {2 [ 1 - \cos(   2   \pi /N ) ]} ~, \\
J_{min} &=& \frac {\hbar \omega} {2 [ 1 - \cos( (N-2) \pi /N ) ]} ~.
\label{eq:jmaxjmin}
\end{eqnarray}
Most crucially, for $J > J_{max}$, it is not possible to resolve the parabolic band bottom that is characteristic of small $\kappa$ from which to extract a meaningful effective mass.
To illustrate the gravity of this restriction, the values of $J_{min}$ and $J_{max}$ so defined have been presented in Table~\ref{tab:cluster} for each of the cluster sizes there considered.
It is clear from these values that cluster size imposes a significant limitation on one's ability to describe polaron band structure over a meaningful interval of $J$.

\section{Effective Mass}

The ground state energy as considered in Section 3 permitted a fair comparison of many different approaches in part because in many cases approximate methodologies are at their ``best'' when applied to the global ground state; excited states almost universally pose greater challenges.
In energy band theory, the minimal excursion beyond the ground state is contained in the effective mass, since in principle the determination of the effective mass requires knowledge of only an infinitesimal excitation to finite $\kappa$.
In this section we compare effective masses as yielded by a number of different approaches.

Our effective mass computations were based on the formula
\begin{equation}
\frac {m^*} {m_0} = \frac {2J} {\frac {\partial^2 E(\kappa)} {{\partial \kappa}^2} |_{\kappa=0}} ~,
\end{equation}
using a discrete representation of the $\kappa$ derivative at the Brillouin zone center.
We note that the $m_0$ used in our calculations was obtained by computing its value using the same discrete differentiation as was applied to compute $m^*$ rather than using the limiting $N \rightarrow \infty$ value.
Not only does this minimize any dependence of the computed effective mass ratio on lattice size, but is technically necessary in order to properly normalize our results; e.g., to exactly recover the limit $\lim_{g \rightarrow 0} m^* /m_0 = 1$.
For all cases presented, $J < J_{max}$ according to (\ref{eq:jmaxjmin}) for lattices of 32 sites, so that our discrete differentiation yields the physically-meaningful value.

\begin{figure}[htb]
\begin{center}
\leavevmode
\epsfxsize = 3.6in
\epsffile{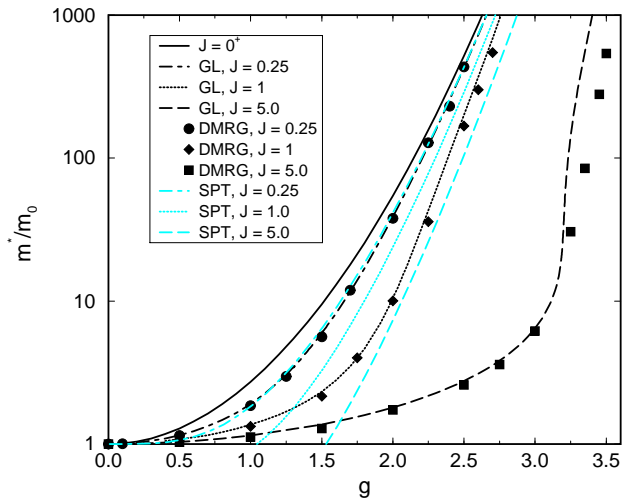}
\end{center}
\caption
{
Effective mass ratio.
Bold curves:  Global-Local results for $J / \hbar \omega =1/4$, $1$, and $5$.
Faint curves:  Second-order strong-coupling perturbation theory (see Eq. \ref{eq:scptmass}).
Points:  Density matrix renormalization group (DMRG) results as kindly provided by E. Jeckelmann \protect\cite{Jeckelmann98,Jeckelmann97b}\protect.
}
\label{fig:white}
\end{figure}

In Figure~\ref{fig:white}, we first consider the simultaneous comparison of our Global-Local effective masses ( actually, $\ln (m^*/m_0)$ ) with comparable DMRG results and with second-order strong-coupling perturbation theory as contained in the relation
\cite{Marsiglio95,Stephan96}
\begin{equation}
\frac {m^*} {m_0} = \frac {e^{g^2}} {1 + (4 J/\hbar \omega) e^{-g^2} f(g^2)} ~,
\label{eq:scptmass}
\end{equation}
for which $f(x)$ has been defined in (\ref{eq:fofx}).

The Global-Local and DMRG results agree quite well, especially considering that these particular DMRG data do not represent actual finite-$\kappa$ calculations, but estimations based on DMRG ground state data \cite{Jeckelmann98,Jeckelmann97b}.
Given the very favorable energy band comparison in Figure~\ref{fig:glvsdmrg}, it is reasonable to expect that a future comparison based on direct excited-state DMRG calculations would show even better agreement (e.g., see Figure~\ref{fig:glvsdmrg}).
What deviations there are between the GL and DMRG results are greatest above the self-trapping transition ($g > g_{ST}$); it is unclear at present whether this is a discrepancy of lasting significance.

On the other hand, comparison of both GL and DMRG effective masses with the results of second-order SCPT are far less favorable.
Agreement is excellent for $J / \hbar \omega =1/4$ and $g > g_{ST}$; however, deviations appear even at this small $J$ value for $g < g_{ST}$.
At $J / \hbar \omega =1$, it is evident that both the GL and DMRG masses asymptote to the SCPT result, but that this convergence of results does not materialize until well above the self-trapping transition ($g \gg g_{ST}$).
At $J / \hbar \omega =5$, the disagreement between the SCPT result and both the GL and DMRG masses is so severe as to render the SCPT estimation useless.
We note that although we have explicitly compared only selected $J$ values, it can be safely inferred that the second-order SCPT mass estimation ceases to be relevant in practical terms by the time $J / \hbar \omega \sim 2$ \cite{Romero98}.

\begin{figure}[htb]
\begin{center}
\leavevmode
\epsfxsize = 3.6in
\epsffile{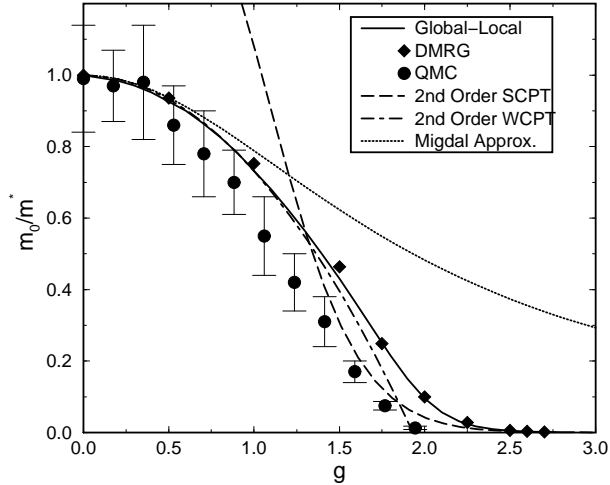}
\end{center}
\caption
{
Inverse effective mass ratio for $J/ \hbar \omega =1$.
Solid curve:  Global-Local result.
Diamond symbols:
DMRG results as kindly provided by E. Jeckelmann
\protect\cite{Jeckelmann98,Jeckelmann97b}\protect.
Bullet symbols:
QMC results as kindly provided by P. Kornilovitch
\protect\cite{Kornilovitch97a,Kornilovitch97b}\protect.
Dotted curve:
Weak-coupling Migdal approximation
(see Eq. \ref{eq:migdalmass}).
Dashed curve:
Second-order strong-coupling perturbation theory
(see Eq. \ref{eq:scptmass}).
Dot-dashed curve:
Second-order weak-coupling perturbation theory
(see Eq. \ref{eq:wcptmass}).
}
\label{fig:emcomp}
\end{figure}

In Figure~\ref{fig:emcomp} we make a more narrow comparison with a broader range of approaches.
Here, we again compare GL and DMRG results, this time including as well \cite{footnote2}:

i) data from direct quantum Monte Carlo calculations of the polaron mass \cite{Kornilovitch97a,Kornilovitch97b},

ii) the second-order SCPT estimate as contained in (\ref{eq:scptmass}),

iii) the second-order WCPT estimate as contained in \cite{Jeckelmann97b}
\begin{eqnarray}
\frac {m_0} {m^*} &=& 1 - \frac {g^2 (1+2J/\hbar \omega )} {(1+4J/\hbar \omega )^{3/2}}
\label{eq:wcptmass}
\end{eqnarray}

and iv) the weak-coupling Migdal estimate as contained in \cite{Migdal58,Capone97}
\begin{equation}
\frac {m^*} {m_0} = 1+ g^2 \frac {(1 + 2J/\hbar \omega )} {(1 + 4J / \hbar \omega )^{3/2}} ~.
\label{eq:migdalmass}
\end{equation}

It is clear that all the presented results except the SCPT are mutually consistent at sufficiently weak coupling; however, divergences between results appear quickly with increasing coupling strength, such that {\it only} the GL, DMRG, and WCPT masses remain mutually consistent to nontrivial coupling.

We note that the agreement between the GL, DMRG, and WCPT masses is actually better than it may appear.
As in Figure~\ref{fig:white}, the DMRG data indicated by diamond symbols is the result of an estimation procedure based on the global DMRG ground state only.
The near-perfect agreement between the GL and DMRG energy bands as shown in Figure~\ref{fig:glvsdmrg} assures that the effective masses computed directly from the complete energy bands are essentially identical.
The same high degree of agreement can be inferred from the comparison of GL and cluster diagonalization bands in Figure~\ref{fig:welleinband}.
The small {\it apparent} discrepancy between the {\it indirect} DMRG estimate and the GL and SCPT results is thus essentially eliminated, showing GL, DMRG, and WCPT to be in essentially complete agreement up to at least $g \sim 1$.

At strong coupling, the GL, DMRG, and SCPT masses are again mutually consistent, while each of the remaining results presented deviates significantly.

\section{Conclusion}

In this paper, we have presented results for the Holstein molecular crystal model in one space dimension as determined by the Global-Local variational method, including complete polaron energy bands, ground state energies, and effective masses.
We have juxtaposed our results with specific comparable results of numerous other methodologies of current interest, including quantum Monte Carlo, cluster diagonalization, dynamical mean field theory, density matrix renormalization group, semiclassical analysis, weak-coupling perturbation theory, and strong-coupling perturbation theory.

Through these comparisons, we have been able to conclude that

i) perturbation theory, while definitive in limits, is clearly superceded by more accurate methods in the intermediate-coupling regime,

ii) semiclassical analysis, while consistent with perturbation theory in the strong-coupling limit, does not improve significantly over perturbation theory away from this limit,

iii) quantum Monte Carlo for the ground state energy is consistent with its best competitors, but owing to limitations of precision is in the present context more confirmatory of other approaches than determinative in itself,

iv) quantum Monte Carlo for the effective mass is consistent with more accurate approaches at weak coupling, but at present is inconsistent with the best estimates at strong coupling,

v) dynamical mean field theory, while exceptionally complete qualitatively, does not deliver superior quantitative accuracy in the regimes considered in this paper.

On the positive side, very favorable comparisons were found between the Global-Local method, the density matrix renormalization group method, and cluster diagonalization methods.
Of these, however, only the GL and DMRG methods could be compared in a broad way.
Both GL and DMRG were found to be consistent with perturbation theory at both weak and strong coupling, and in the weak-coupling case, at least, consistent to higher coupling strengths than any other method.
Where GL and DMRG depart from perturbation theoretic results, in the intermediate coupling regime, GL and DMRG remain mutually consistent to an impressive degree, and where direct comparison has been possible, consistent with cluster diagonalization.

We take these results as a whole to confirm the Global-Local variational method as being highly accurate over a wide range of the polaron parameter space, from the non-adiabatic limit to the extremes of high adiabaticity, from weak coupling through intermediate coupling to strong coupling.

In succeeding works \cite{Romero98b,Romero98c,Romero98}, we shall present a comprehensive analysis of polaron structure and properties as determined in the Global-Local method over essentially the whole of the polaron parameter space.

\section*{Acknowledgment}

This work was supported in part by the U.S. Department of Energy under Grant No. DE-FG03-86ER13606.
The authors gratefully acknowledge the cooperation of
H. De Raedt and A. Lagendijk \cite{DeRaedt83,DeRaedt97},
P. E. Kornilovitch and E. R. Pike \cite{Kornilovitch97a,Kornilovitch97b},
A. S. Alexandrov, V. V. Kabanov, and D. E. Ray \cite{Alexandrov94a,Kabanov97},
G. Wellein and H. Fehske \cite{Wellein97a,Wellein97b},
E. Jeckelmann and S. White \cite{Jeckelmann98,Jeckelmann97b}, and
G. Kalosakas, S. Aubry, and G. P. Tsironis \cite{Kalosakas97a,Kalosakas97b},
for providing numerical values of data used in parts of this paper.

\bibliography{../../Bibliography/theory,../../Bibliography/books,../../Bibliography/experiment,../../Bibliography/temporary}

\newpage


\newpage


\newpage

\begin{table*}[t]
\centering
\vskip 2mm
\begin{tabular}{|l|l|l|} \hline
Value $[ \hbar \omega ]$ & Type of Method & Source \\ \hline \hline
- 1.73575 $\dagger$ & Non-adiabatic small polaron & Holstein \cite{Holstein59b}, Eq. \ref{eq:smpolgrnd}\\ \hline
- 1.97 & QMC $N = 32$ $\beta = 1$ $m = 10$ & De Raedt and Lagendijk \cite{DeRaedt83,DeRaedt97} \\ \hline
- 2.00000 $\dagger$ & Adiabatic strong coupling & Gogolin \cite{Gogolin82}, Eq. \ref{eq:gogolin}\\ \hline
- 2.08614 & Semiclassical variation & Kalosakas, Aubry, and Tsironis \cite{Kalosakas97a,Kalosakas97b} \\ \hline
- 2.35 & QMC $N = 32$ $\beta = 5$ $m = 32$ & De Raedt and Lagendijk \cite{DeRaedt83,DeRaedt97} \\ \hline
- 2.40 & Dynamical Mean Field & Ciuchi, et al. \cite{Ciuchi97} \\ \hline
- 2.44721 & $2^{nd}$ order WCPT & Eq. \ref{eq:wcptgrnd} \\ \hline
- 2.45611 & Merrifield variation & Zhao, Brown, and Lindenberg \cite{Zhao97a} \\ \hline
- 2.46 & QMC $N = 32$ $\beta = 20$ $m = 256$ & De Raedt \cite{DeRaedt97} \\ \hline
- 2.46869 & Toyozawa variation & Zhao, Brown, and Lindenberg \cite{Zhao97b} \\ \hline
- 2.46931 & Global-Local variation & Figure~\ref{fig:grndj}, this paper \\ \hline
- 2.46968 & DMRG $N = 32$ & Jeckelmann and White \cite{Jeckelmann98,Jeckelmann97b} \\ \hline \hline
- 2.471 & Cluster diagonalization, $N = 6$ & Alexandrov, et al. \cite{Alexandrov94a,Kabanov97,footnote1} \\ \hline
- 3.08959 $\dagger$ & $2^{nd}$ order SCPT & Marsiglio \cite{Marsiglio95}, Stephan \cite{Stephan96}, Eq. \ref{eq:scptgrnd}\\ \hline
\end{tabular}
\end{table*}

\clearpage

\begin{table*}
\centering
\vskip 2mm
\begin{tabular}{|l|l|l|} \hline
Value $[ \hbar \omega ]$ & Type of Method & Source \\ \hline \hline
- 2.27067 $\dagger$ & Non-adiabatic small polaron & Holstein \cite{Holstein59b}, Eq. \ref{eq:smpolgrnd}\\ \hline
- 2.50000 $\dagger$ & Adiabatic strong coupling & Gogolin \cite{Gogolin82}, Eq. \ref{eq:gogolin}\\ \hline
- 2.51815 & Semiclassical variation & Kalosakas, Aubry, and Tsironis \cite{Kalosakas97a,Kalosakas97b} \\ \hline
- 2.73 & QMC $N = 32$ $\beta = 1$ $m = 10$ & De Raedt and Lagendijk \cite{DeRaedt83,DeRaedt97} \\ \hline
- 2.86 & QMC $N = 32$ $\beta = 5$ $m = 32$ & De Raedt and Lagendijk \cite{DeRaedt83,DeRaedt97} \\ \hline
- 2.89 & Dynamical Mean Field & Ciuchi, et al. \cite{Ciuchi97} \\ \hline
- 2.89442 & $2^{nd}$ order WCPT & Eq. \ref{eq:wcptgrnd} \\ \hline
- 2.93301 & Merrifield variation & this paper \\ \hline
- 2.99172 & Toyozawa variation & this paper \\ \hline
- 2.99802 & Global-Local variation & Figure~\ref{fig:grndj}, this paper \\ \hline
- 2.99883 & DMRG $N = 32$ & Jeckelmann and White \cite{Jeckelmann98,Jeckelmann97b} \\ \hline \hline
- 3.000 & Cluster diagonalization, $N = 6$ & Alexandrov, et al. \cite{Alexandrov94a,Kabanov97,footnote1} \\ \hline
- 3.00 & QMC $N = 32$ $\beta = 20$ $m = 256$ & De Raedt \cite{DeRaedt97} \\ \hline
- 3.05279 $\dagger$ & $2^{nd}$ order SCPT & Marsiglio \cite{Marsiglio95}, Stephan \cite{Stephan96}, Eq. \ref{eq:scptgrnd}\\ \hline
\end{tabular}
\end{table*}

\clearpage

\begin{table*}[t]
\centering
\vskip 2mm
\begin{tabular}{|r|c|c|r|r|r|} \hline
Cluster & $J_{min} / \hbar \omega $ & $J_{max} / \hbar \omega $ & $D_{tot}, M=10$ & $D_{tot}, M=16$ & $D_{tot}, M=32$ \\ \hline \hline
2 sites & no value & no value & $132$ & $306$ & $1122$ \\ \hline
4 sites & $0.5000$ & $0.500$ & $4.0 \times 10^3$ & $1.9 \times 10^4$ & $2.4 \times 10^5$ \\ \hline
6 sites & $0.3333$ & $1.000$ & $4.8 \times 10^4$ & $4.5 \times 10^5$ & $1.7 \times 10^7$ \\ \hline
10 sites & $0.2764$ & $2.618$ & $1.8 \times 10^6$ & $5.3 \times 10^7$ & $1.5 \times 10^{10}$ \\ \hline
16 sites & $0.2599$ & $6.569$ & $8.5 \times 10^7$ & $9.6 \times 10^9$ & $3.6 \times 10^{13}$ \\ \hline
20 sites & $0.2563$ & $10.22$ & $6.0 \times 10^8$ & $1.5 \times 10^{11}$ & $2.5 \times 10^{15}$ \\ \hline
32 sites & $0.2524$ & $26.02$ & $4.7 \times 10^{10}$ & $7.2 \times 10^{13}$ & $5.9 \times 10^{19}$ \\ \hline
\end{tabular}
\end{table*}

\end{document}